\documentstyle[preprint,prb,aps]{revtex}

\begin{document}
\draft
\title{Exciton optical absorption in self-similar aperiodic lattices}

\author{Enrique Maci\'{a}$^*$ and
Francisco Dom\'{\i}nguez-Adame$^{\dag}$}

\address{Departamento de F\'{\i}sica de Materiales,
Facultad de F\'{\i}sicas, Universidad Complutense,
E-28040 Madrid, Spain}

\maketitle

\begin{abstract}

Exciton optical absorption in self-similar aperiodic one-dimensional
systems is considered, focusing our attention on Thue-Morse and
Fibonacci lattices as canonical examples.  The absorption line shape is
evaluated by solving the microscopic equations of motion of the
Frenkel-exciton problem on the lattice, in which on-site energies take
on two values, according to the Thue-Morse or Fibonacci sequences.
Results are compared to those obtained in random lattices with the same
stechiometry and size.  We find that aperiodic order causes the
occurrence of well-defined characteristic features in the absorption
spectra which clearly differ from the case of random systems, indicating
a most peculiar exciton dynamics.  We successfully explain the obtained
spectra in terms of the two-center problem.  This allows us to establish
the origin of all the absorption lines by considering the self-similar
aperiodic lattices as composed of two-center blocks, within the same
spirit of the renormalization group ideas.

\end{abstract}


\narrowtext

\section{Introduction}

During the last few years the notion of {\em aperiodic order} has
progressively emerged to gain a proper understanding of new physical
systems.  Since the remarkable discovery of the quasicrystalline phase
\cite{Shechtman} and the technical advances in submicron physics for the
fabrication of semiconductor superlattices arranged according to the
(quasiperiodic) Fibonacci \cite{Merlin} and Thue-Morse
\cite{Merlin2,Axel} sequences, much work has been devoted to the study
of systems whose structural order is described by means of deterministic
substitution sequences, \cite{Que} leading to self-similar aperiodic
lattices.  The interest in exploring the physical properties of
elementary excitations in one-dimensional (1D) aperiodic systems,
including Fibonacci, Thue-Morse or Rudin-Shapiro lattices and their
generalizations, goes beyond a formal theoretical analysis of systems
deserving a simpler mathematical treatment than three-dimensional ones.
In fact, it is actually well known that aperiodic order gives rise to
novel properties which are completely absent in both periodic
(crystalline) and random (amorphous) 1D systems.  In this way, aperiodic
systems exhibit highly fragmented electron \cite{Severin,Angel,Oh} and
phonon \cite{Kohmoto,Zhong} spectra that are Cantor sets determining the
existence of critical states.  These exotic electronic spectra strongly
influence electron propagation, being somewhat intermediate between
ballistic and difussive, which gives rise to unusual behavior of the dc
conductance at finite temperature. \cite{PRBFIBO} Although these
striking results were initially obtained for tight-binding and
Kronig-Penney model Hamiltonians, we have recently shown that peculiar
electronic transport properties must be also expected in more realistic
systems. \cite{PLAFIBO}

Moreover, it has also been realized that systems ordered according to
the Fibonacci sequence exhibit some characteristic properties which are
not shared by other self-similar aperiodic arrangements.  In particular,
from studies concerning the electronic spectrum structure, \cite{Ryu}
Landauer resistance, \cite{Huang,Roy} and phonon spectrum properties,
\cite{Tamura} some authors have claimed that the kind of order
associated to the Thue-Morse sequence must be considered as intermediate
between the quasiperiodic order displayed by Fibonacci systems and the
usual periodic order.  We believe that this result is not surprising.
In fact, the Fourier spectrum of Fibonacci lattices is pure point, hence
indicating the existence of long-range order associated to the
quasiperiodic nature of the underlying lattice, whereas the Fourier
spectrum of Thue-Morse lattices is singular continuous. \cite{Kolar}
Nevertheless, we feel that, in some of the above mentioned works, the
criteria introduced to determine the {\em degree of periodicity}
associated to a given self-similar lattice are somewhat vague and could
lead to possible misinterpretations. \cite{Bovier} Therefore, to attain
a deeper insight into the nature of the order displayed by different
kinds of aperiodic structures, it seems convenient to investigate
transport properties different from those usually considered (electron
propagation, phonon dynamics).

In this regard, optical properties of aperiodic lattices have received
much less attention and, to our knowledge, most of the work has been
restricted to the study of optical phenomena in Fibonacci superlattices.
\cite{Tuet,Munzar} In this work we investigate the optical absorption
spectra of two different kinds of self-similar aperiodic systems, namely
the Fibonacci lattice (FL) and the Thue-Morse lattice (TML), and compare
them with the optical spectra characteristic of both binary random and
periodic related systems.  To this end, we make use of a general
treatment which allows us to study the dynamics of Frenkel-excitons in
these lattices, solve the microscopic equations of motion and find the
optical absorption spectra.  This study is inspired in our previous work
showing that short-range correlated disorder has profound effects on
trapping \cite{PRBTRAP} and optical properties \cite{PRBOPT} of
Frenkel-exciton systems.  These results lead, in a natural way, to the
question as to whether long-range aperiodic order modifies exciton
dynamics in comparison to long-range disorder effects.  The main
conclusions of this work are twofold.  First, we show that both FLs and
TMLs exhibit optical absorption spectra quite different from those
obtained in random and periodic lattices.  Therefore, optical spectra
can be used to characterize experimentally the occurrence of aperiodic
order in the sample.  Second, we show that optical spectra are able to
discriminate also the particular kind of aperiodic order present in the
system.  Hence, the analysis of optical absorption spectra appears as an
excellent {\em diagnostic tool} to characterize the structural order
from an experimental point of view.

The remaining of the paper is organized as follows.  In Sec.~II we
describe our model and the different self-similar aperiodic arrangements
we are going to investigate, and we show how optical spectra can be
numerically obtained.  In Sec.~III we give a detailed account of the
main lines appearing in the spectra, and we compare them with spectra of
random and periodic systems.  Section IV is devoted to find a
relationship between those features and the underlying lattice topology
by means of the so-called two-center problem, guided by renormalization
group ideas.  Then, the origin of the main lines appearing in the
spectra is explained in Sec.~V on the basis of the two-center model.
Section VI concludes the paper with some general remarks on the physical
implications and possible extensions of our results.

\section{Model}

We consider a system of $N$ optically active centers, occupying
positions in a regular 1D lattice with spacing unity.  For our present
purposes we neglect all thermal degrees of freedom, and thus we omit
electron-phonon coupling and local lattice distortions.  Therefore, the
effective Hamiltonian that describes the Frenkel-exciton problem can be
written in the well-known tight-binding form with nearest-neighbor
interactions as follows (we use units such that $\hbar=1$)
\begin{equation}
{\cal H}= \sum_{k}\> V_{k} a_{k}^{\dag}a_{k} +
T\sum_{k}\>(a_{k}^{\dag}a_{k+1}+a_{k+1}^{\dag}a_{k}).
\label{perfectH}
\end{equation}
Here $a_{k}^{\dag}$ and $a_{k}$ are Bose operators creating and
annihilating an electronic excitation of energy $V_k$ at site $k$,
respectively. $T$ is the nearest-neighbor coupling, which is assumed to
be constant in the whole lattice.  In what follows we consider that
$V_k$ can only take on two values, $V_A$ and $V_B$, and we shall arrange
them either aperiodically, according to the Thue-Morse and Fibonacci
sequences, or randomly.  For convenience we define $c_A$  ($c_B=1-c_A$)
as the ratio between the number of sites $A$ ($B$) and the total number
of sites $N$ in the lattice.

TML and FL are canonical examples of deterministic and aperiodically
ordered systems, and they can be generated by the following substitution
rules: $A\rightarrow AB$, $B\rightarrow BA$ for the Thue-Morse sequence
and $A\rightarrow AB$, $B\rightarrow A$ for the Fibonacci one.  In this
way, finite and self-similar aperiodic lattices are obtained by $n$
successive applications of the substitution rule.  The {\em n\/}th
generation lattice will have $2^n$ elements in the TML and $F_n$
elements for the FL, where $F_n$ denotes a Fibonacci number.  Such
numbers are generated from the recurrence law $F_n=F_{n-1}+F_{n-2}$
starting with $F_0=F_1=1$; as $n$ increases the ratio $F_{n-1}/F_n$
converges toward $\tau=(\sqrt{5}-1)/2=0.618\ldots$, an irrational number
which is known as the inverse golden mean.  Therefore, the on-site
excitation energies are arranged according to the sequence
$V_A\,V_B\,V_B\,V_A\,V_B\,V_A\,V_A\,V_B\ldots$ in the TML and
$V_A\,V_B\,V_A\,V_A\,V_B\,V_A\,V_B\,V_A\ldots$ in the FL. The values of
$c_A$ and $c_B$ are strictly equal to $0.5$ for any generation of the
TML. On the contrary, the values of $c_A$ and $c_B$ depend on the
particular generation of the FL, but for large systems one has
$c_A\sim\tau$ and $c_B\sim 1-\tau$.  Finally, it is worth noticing that
$B$-centers appear isolated in FLs. This is an important fact in order
to explain the results we will present later.

Having presented our model we now briefly describe the method we have
used to calculate the absorption spectra.  The line shape $I(E)$ of
an optical-absorption process in which a single exciton is created in a
lattice with $N$ sites can be obtained as follows. \cite{Huber} Let us
introduce a set of correlation functions
\begin{equation}
G_{k}(t)=\sum_{j}\> \langle 0|a_{k}(t)a_{j}^{\dag}|0\rangle,
\label{G}
\end{equation}
where $|0\rangle$ denotes the exciton vacuum state and $a_{k}(t)= \exp
(i{\cal H}t) a_{k} \exp (-i{\cal H}t)$ is the annihilation operator in
the Heisenberg representation.  The function $G_{k}(t)$ obeys the
equation of motion
\begin{equation}
i{d\over dt} G_{k}(t) = \sum_{j}\> H_{kj} G_{j}(t),
\label{motion}
\end{equation}
with the initial condition $G_{k}(0)=1$.  The diagonal elements of the
tridiagonal matrix $H_{kj}$ are $V_{k}$ whereas off-diagonal elements
are simply given by $T$.  The micrsocopic equation of motion is a
discrete Schr\"odinger-like equation on a lattice and standard numerical
techniques may be applied to obtain the solution.  Once these equations
of motion are solved, the line shape is found from the following
expression
\begin{equation}
I(E)=-\,{2\over \pi N} \int_0^\infty\> dt\, e^{-\alpha t} \sin (Et)
\,\mbox{\rm Im} \left( \sum_k\> G_k(t) \right),
\label{line}
\end{equation}
where the factor $\exp (-\alpha t)$ takes into account the broadening
due to the Lorentzian instrumental resolution function of width
$\alpha$.

\section{Results}

We have solved numerically the equation of motion (\ref{motion}) using
an implicit (Crank-Nicholson) integration scheme.  In the remainder of
the paper, energy will be measured in units of $T$ whereas time will be
expressed in units of $T^{-1}$.  Aperiodic lattices are generated using
the inflation rules discussed above.  We have checked that the main
features of the spectra are independent of the system size.  Henceafter
we will fix $N=2^{11}=2048$ for the TML and $N=F_{16}=1597$ for the FL
as representative values.  In addition, standard random generators are
used to obtain disordered lattices with the required size $N$ and value
of $c_B$ ($N=2048$, $c_B=0.5$ to compare with TML and $N=1597$ and
$c_B=0.382$ to compare with FL). In order to minimize end effects,
spatial periodic boundary conditions are introduced in all cases.  Once
the functions $G_k(t)$ are known, the line shape $I(E)$ is evaluated by
means of (\ref{line}).  Since we are mainly interested in characterizing
the effects due to aperiodic order as compared to randomness rather
than in a detailed discussion of the optical absorption process, we will
fix the values of $V_A$, $V_B$ and $T$, focusing our attention on the
comparison between different types of arrangements of optical centers.
Furthermore, in order to facilitate the comparison with our previous
work, we have set $V_A=4$, $V_B=10$ and $T=-1$ henceafter.  The width of
the instrumental resolution was $\alpha=0.5$.  The maximum integration
time and the integration time step were $16$ and $8\times 10^{-3}$,
respectively; larger maximum integration times or smaller time steps led
to the same general results.

{}For the sake of clarity let us consider, in the first place, the
typical spectra associated to both pure $A$ and pure $B$ lattices
corresponding to periodic cases.  In the pure $A$ lattice the spectrum
is a single Lorentzian line centered at $E_{\scriptstyle pure}^{A} =
V_A+2T$, which with our choice of parameters is $E_{\scriptstyle
pure}^{A}=2.0$.  When $B$-centers are introduced at random in the
lattice, a broadening of this main line is observed accompanied by a
shift of its position towards higher energies.  In random systems both
the broadening and the shift increase on increasing defect
concentration, \cite{PRBOPT} in agreement with the average-T-matrix
approximation (ATA). \cite{Huber} A similar behavior takes place in the
pure $B$ lattice when $A$-centers are introduced with the main
difference that, in this case, the single Lorentzian absorption line is
originally located at $E_{\scriptstyle pure}^{B}=V_B+2T=8.0$.  However,
ATA is no longer valid to determine optical spectra in TML and FL due to
the long-range correlation induced by aperiodic order, as we will
discuss later.

Keeping these general results in mind we now proceed to discuss the main
features of the spectra obtained in aperiodic systems.  We shall start
with the TML and compare it to a typical random lattice.  The obtained
results are shown in Fig.~\ref{fig1}, where all the spectra have the
same area.  From a close inspection of this figure several conclusions
can be drawn.  First of all, we observe the occurrence of a strong line
centered at $E=2.9$ in the TML. This line is accompanied by a small
shoulder at around $E=3.8$ (the position of the shoulder has been
obtained using two Lorentzian functions to fit data in the energy range
from $0$ up to $6$).  Moreover, two satellites appear in the high-energy
region of the spectrum at about $E=9.0$ and $E=10.2$.  On the other
hand, concerning the random lattice, we note that the main absorption
line is centered at about $E=2.6$, closer to the position corresponding
to the single line in the pure $A$ lattice although the small shoulder
remains at about $E=3.8$.  In addition, the intensity of the overall
absorption features in the energy range $0 \leq E \leq 6$ is smaller
than those corresponding to the TML. To conclude, we observe that the
random lattice also presents a characteristic pair of satellites at the
high-energy region of the spectrum.  One of them is centered at
$E=10.2$, as occurs in the TML, but the other is found at $E=8.2$.
Finally, the satellite at $E=9.0$, clearly observed in the TML, appears
as an almost unnoticeable shoulder causing the asymmetry of the line at
$E=8.2$.

Let us now turn to the FL and compare it to a random lattice with the
same size $N$ and $c_B$.  Results are shown in Fig.~\ref{fig2}.  The FL
presents two clearly distinct lines.  In the low-energy range we observe
a main absorption line centered $E=2.9$ embodying an almost unobservable
shoulder at $E=3.8$ meanwhile, in the high-energy region of the
spectrum, a single satellite at $E=10.2$ is observed.  On the other
side, the spectrum associated to the random lattice shows a main
absorption line at about $E=2.3$ along with a smaller shoulder at about
$E=3.8$ in the low-energy region whereas, at higher energies, two broad
satellites are clearly observable at $E=9.0$ and $E=10.2$.

By comparing Figs.~\ref{fig1} and \ref{fig2} we are led to the
conclusion that, on the basis of the observed optical absorption
spectra, very significant differences exist, not only between aperiodic
systems and the corresponding random ones, but also between the two
realizations of aperiodic order we have considered.  In fact, on the one
hand, besides the small shoulder above mentioned, there are only two
distinct lines in the absorption spectrum of the FL whereas three
different lines are clearly observable in the spectrum corresponding to
the TML. On the other hand, the absorption lines at $E=2.9$ and $E=10.2$
are more intense in the FL spectrum than in the TML one.  Finally, the
satellite peak at $E=9.0$, clearly visible in the TML spectrum is not
observed at all in the FL case.  In the following Section we explain the
origin of these characteristic features in terms of short-range quantum
effects.

\section{The two-center problem}

One of the most remarkable aspects of the electronic spectra in 1D
aperiodic systems is their highly fragmented nature which corresponds to
a Cantor-like set with zero Lebesgue measure in the thermodynamical
limit.  The fragmentation pattern of the spectra varies on increasing
the generation of the lattice and their detailed structure is mainly
determined by short-range effects.  This point was earlier suggested by
means of the real space renormalization group, where the number of
energy levels appearing at the first stage of the renormalization
process determines the number of main clusters in the spectra.
\cite{Niu,Liu} This number of levels depends on the adopted blocking
scheme which, for a binary system within the weak bound approach,
usually decouples the original lattice in a series of single ($A$, $B$)
or double ($AA$, $AB$, $BB$) constituent elements.  The procedure just
outlined justifies the purported asymptotic stability of the electronic
spectra of electronic Fibonacci systems. \cite{PRBFIBO} Furthermore, we
have recently shown that the great success of the renormalization group
scheme can be directly traced back to the fact that systems generated
from the application of a substitution sequence encode more information,
in the Shannon sense, than classical periodic systems. \cite{PRE} As a
consequence, from the assumption that the lattice topology must have
profound influences on the exciton dynamics, it seems natural to extend
the main ideas inspiring the renormalization procedure to account for
the origin of the different lines appearing in the optical spectra of
aperiodic systems.

To this end, we shall consider the two-center problem describing the
optical absorption spectrum of two isolated but coupled sites, labelled
$1$ and $2$.
\begin{mathletters}
\label{two}
\begin{eqnarray}
i{d\over dt} G_{1}(t) &=& V_1G_{1}(t)+T G_{2}(t), \label{twoa} \\
i{d\over dt} G_{2}(t) &=& V_2G_{2}(t)+T G_{1}(t). \label{twob}
\end{eqnarray}
\end{mathletters}
Solving these equations exactly with the initial conditions $G_1(0) =
G_2(0) = 1$ and inserting the result in (\ref{line}) one obtains (we
neglect the instrumental resolution for simplicity)
\begin{equation}
I(E)= I_{+}\delta (E-E_{+})+ I_{-}\delta (E-E_{-}),
\label{spectratwo}
\end{equation}
where
\begin{equation}
I_{\pm}={1\over 2} \mp {1\over 2}\left[1+\left(
{V_1-V_2\over 2T} \right)^2\right]^{-1/2}
\label{i}
\end{equation}
and
\begin{equation}
E_{\pm}={V_1+V_2\over 2}\mp T\sqrt{1+\left( {V_1-V_2\over 2T}\right)^2}.
\label{e}
\end{equation}

As expected, the optical absorption spectrum of the two-center problem
presents two well-defined lines.  From the Eqs.~(\ref{i}) and (\ref{e})
we arrive at the following possible situations.  If $V_1$=$V_2$, the
intensity $I_{+}$ vanishes so that spectrum exhibits a single line.
Depending on the nature of the centers this line will be centered either
at $E_{-}^{AA}=V_A+T=3.0$ ($AA$ pairs) or $E_{-}^{BB}=V_B+T=9.0$ ($BB$
pairs).  On the contrary, if the on-site excitation energies are
different (say $V_1=V_A$ and $V_2=V_B$) the optical spectrum presents
two components centered at $E_{\pm}^{AB}=(V_A+V_B)/2 \mp T \sqrt{1+(V_A
- V_B)^2/4T^2}=7\pm \sqrt{10}$ so that, in our units, $E_{-}^{AB}\sim
3.8$ and $E_{+}^{AB}\sim 10.2$.

Therefore, with the aid of the two-center model we can uniquely assign
specific absorption lines to each of the pairs in which our original
lattice can be decomposed, according to the renormalization group ideas
mentioned above.  In this sense, the signatures of $AA$ and $BB$ pairs
are single lines located at $E_{-}^{AA}=3.0$ and $E_{-}^{BB}=9.0$.  In
addition, $AB$ or $BA$ pairs can be associated, irrespectively, to the
simultaneous presence of two characteristic lines in the spectrum,
centered at $E_{-}^{AB} \simeq 3.8$ and $E_{+}^{AB} \simeq 10.2$.
According with these precise assignments, the origin of the main lines
and satellites appearing in the absorption spectra of both FLs and TMLs
can be unequivocally established.

\section{Discussion}

In this section we explain the origin of the lines appearing in
Figs.~\ref{fig1} and \ref{fig2} making use of the two-center results.
For convenience, we shall discuss both kinds of aperiodic lattices
separately.

\subsection{Thue-Morse absorption spectrum}

Let us focus our attention on Fig.~\ref{fig1}.  The main line centered
at $E=2.9$ is very close to the characteristic line $E_{-}^{AA}=3.0$
associated to the $AA$ pair, hence strongly suggesting the possible
origin of this absorption line.  At this point it is important to note
that the ATA approach cannot account for the presence of this line,
since the shift of the main Lorentzian at $E_{\scriptstyle
pure}^{A}=2.0$ in the pure $A$ lattice due to the presence of
$B$-centers in a concentration $c_{B}=0.5$ amounts to only $0.6$ units.
This value is more than $30 \%$ lower than that obtained in the TML
spectrum.  Thus, it becomes clear that the main absorption line observed
in the TML is not simply the $E_{\scriptstyle pure}^{A}=2.0$ line
shifted by the presence of $B$-centers, as occurs in the random lattice.
This result suggests that the aperiodic order displayed by the TML has
profound effects on the resulting exciton dynamics, which in turn
manifest in the optical spectra.  To find a heuristic explanation of the
different exciton dynamics in TML, we would like to draw the attention
on the fact that the $A$-centers can appear only isolated or grouped in
pairs in the TML, but never forming larger groups, as it could be the
case in random lattices.  In fact, the presence of these larger
segments, which behave locally as pure $A$ lattice segments having
$B$-centers at the ends, is what explains the shift of $E_{\scriptstyle
pure}^{A}=2.0$ line towards higher energies within the framework of ATA.
Therefore, the absence of large groups of $A$-centers in the TML, along
with the relative abundance of $AA$ pairs instead, causes the occurrence
of a noticeable and high peak at $\sim 3.0$.  Similar reasoning explains
the absence of the $E_{\scriptstyle pure}^{B}=8.0$ line in the TML
spectrum, whereas such a line is clearly seen in the corresponding
random lattice spectrum, shifted to $E=8.2$ due to the presence of
$A$-centers.  Furthermore, the marked satellite at $E_{-}^{BB}=9.0$ must
be associated to the presence of many $BB$ pairs in the TML. Finally,
the characteristic absorption satellites associated to $AB$ pairs are
revealed as a shoulder ($E_{-}^{AB}=3.8$) at the high-energy side of the
main line and as a absorption line at $E_{+}^{AB}=10.2$.  In order to
further confirm this identification we have made use of Lorentzian
fitting of data to evaluate the ratio between the relative intensity of
lines at $E_{-}^{AB}=3.8$ and $E_{+}^{AB}=10.2$, which is found to be
$I_{-}/I_{+} \simeq 2.04$.  This value agrees rather well with the
theoretical estimation $I_{-}/I_{+}= (\sqrt{10}+1) / (\sqrt{10}-1)\simeq
1.92$ obtained from Eq.~(\ref{i}).

\subsection{Fibonacci absorption spectrum}

Now we turn our attention on the Fig.~\ref{fig2}.  Once again the
contribution due to $AA$ pairs (main peak at $2.9$) and $AB$ pairs
(small shoulder at $\simeq 3.8$ and satellite peak at $10.2$) are
clearly seen in the absorption spectrum, hence supporting the
convenience of our the two-center description.  Moreover, one of the
most remarkable characteristic of this spectrum, as compared to that
corresponding to the TML, is the dramatic absence of the $E_{-}^{BB} =
9.0$ line.  In fact, according to our previous discussion such a line
comes from the contribution of $BB$ pairs but, as it is well known, such
pairs are forbidden in the FL. We feel this is a very significant result
since it further confirms the correctnes of our interpretation about the
origins of the different lines appearing in the spectra and, at the same
time, allows for an easy and confident differentiaton between different
kinds of aperiodic self-similar lattices from an experimental point of
view.

\section{Conclusions}

In summary, we have studied the absorption spectra corresponding to the
Frenkel-exciton Hamiltonian on self-similar aperiodic systems described
by the Thue-Morse and Fibonacci sequences.  By comparing the obtained
spectra with those corresponding to random lattices we conclude that FLs
and TMLs exhibit characteristic absorption spectra, different in many
aspects from those of binary random lattices with the same stechiometry,
and that certain spectral lines can be used to characterize the
aperiodic order associated to FLs from that related to TMLs.  On the
other side, from the viewpoint of physical applications, we have
obtained analytical expressions which explain our spectra and relate
microscopic system parameters like on-site excitation energies, to
experimental data like position and strengths of the lines.  This
relationship surely should facilitate future experimental work on
optical properties of quasicrystalline solids.

Our treatment allows us to introduce, in a rather straightforward and
natural way, concepts inspired in renormalizaton group techniques, which
have accomplished a great success in describing the electronic spectra
of aperiodic systems.  On the light of the obtained results and previous
discussions, we think that the question as to whether Thue-Morse systems
are more or less periodic than Fibonacci ones, a controversy which has
raised some debate during the last few years, is still ill posed.  In
our opinion, both Thue-Morse and Fibonacci systems display a new kind of
order, namely self-similar aperiodic order, which has its own
peculiarities, and cannot be compared with periodically ordered systems
in a simple way.  In fact, we have recently shown that self-similar
aperiodically ordered systems are able to encode more information, in
the Shannon sense, than usual periodic ones \cite{PRE}, thus opening a
new way to aswer this question.  This line of reasonings may lead to a
novel vision on the concept of order.  Rather than to think into
different kinds of order, classified into separate categories which are
{\em compared} in a quantitative way (in the sense above mentioned of a
particular category to be less random or more periodic than any other
one), maybe more fruitful to think into different {\em hierarchies of
order}.  This perspective, which is inspired into the mathematical
relationships between periodic, quasiperiodic and almost periodic
functions, might be of interest to those researchers working on this
field.

\acknowledgments

The authors thank A.\ S\'{a}nchez for a critical reading of the
manuscript.  This work is partially supported by Universidad Complutense
through project PR161/93-4811.

\begin{figure}
\caption{Absorption spectra for a 1D Thue-Morse lattice (solid line) and
a random lattice (dashed line).  In both cases the system size is
$N=2048$ and the concentration of $B$-centers is $c_B=0.5$.}
\label{fig1}
\end{figure}

\begin{figure}
\caption{Absorption spectra for a 1D Fibonacci lattice (solid line) and
a random lattice (dashed line).  In both cases the system size is
$N=1597$ and the concentration of $B$-centers is $c_B=0.382$.}
\label{fig2}
\end{figure}

\end{document}